\documentclass[12pt]{article}
\usepackage{graphicx}
\usepackage{color}

\def\hybrid{\topmargin 0pt      \oddsidemargin 0pt
        \headheight 0pt \headsep 0pt
       \voffset-1cm
        \textwidth 6.25in       
       \textheight 9.5in       
        \marginparwidth 0.0in
        \parskip 5pt plus 1pt   \jot = 1.5ex}
\catcode`\@=11
\def\marginnote#1{}

\newcount\hour
\newcount\minute
\newtoks\amorpm
\hour=\time\divide\hour by60
\minute=\time{\multiply\hour by60 \global\advance\minute by-\hour}
\edef\standardtime{{\ifnum\hour<12 \global\amorpm={am}%
        \else\global\amorpm={pm}\advance\hour by-12 \fi
        \ifnum\hour=0 \hour=12 \fi
        \number\hour:\ifnum\minute<10 0\fi\number\minute\the\amorpm}}
\edef\militarytime{\number\hour:\ifnum\minute<10 0\fi\number\minute}

\def\draftlabel#1{{\@bsphack\if@filesw {\let\thepage\relax
   \xdef\@gtempa{\write\@auxout{\string
      \newlabel{#1}{{\@currentlabel}{\thepage}}}}}\@gtempa
   \if@nobreak \ifvmode\nobreak\fi\fi\fi\@esphack}
        \gdef\@eqnlabel{#1}}
\def\@eqnlabel{}
\def\@vacuum{}
\def\draftmarginnote#1{\marginpar{\raggedright\scriptsize\tt#1}}

\def\draftlabel#1{{\@bsphack\if@filesw {\let\thepage\relax
   \xdef\@gtempa{\write\@auxout{\string
      \newlabel{#1}{{\@currentlabel}{\thepage}}}}}\@gtempa
   \if@nobreak \ifvmode\nobreak\fi\fi\fi\@esphack}
        \gdef\@eqnlabel{#1}}
\def\@eqnlabel{}
\def\@vacuum{}
\def\draftmarginnote#1{\marginpar{\raggedright\scriptsize\tt#1}}

\def\draft{\oddsidemargin -.5truein
        \def\@oddfoot{\sl preliminary draft \hfil
        \rm\thepage\hfil\sl\today\quad\militarytime}
        \let\@evenfoot\@oddfoot \overfullrule 3pt
        \let\label=\draftlabel
        \let\marginnote=\draftmarginnote
   \def\@eqnnum{(\theequation)\rlap{\kern\marginparsep\tt\@eqnlabel}%
\global\let\@eqnlabel\@vacuum}  }

\def\numberbysection{\@addtoreset{equation}{section}
        \def\theequation{\thesection.\arabic{equation}}}

\def\underline#1{\relax\ifmmode\@@underline#1\else
        $\@@underline{\hbox{#1}}$\relax\fi}

\def\titlepage{\@restonecolfalse\if@twocolumn\@restonecoltrue\onecolumn
     \else \newpage \fi \thispagestyle{empty}\c@page\z@
        \def\thefootnote{\fnsymbol{footnote}} }

\def\endtitlepage{\if@restonecol\twocolumn \else  \fi
        \def\thefootnote{\arabic{footnote}}
        \setcounter{footnote}{0}}  
\relax

\hybrid

\newfont{\Bbb}{msbm10 scaled 1\@ptsize00}
\newfont{\Bbbb}{msbm7 scaled 1\@ptsize00}

\newcommand{\DDD}{\raise-1pt\hbox{$\mbox{\Bbbb D}$}}

\newcommand{\UUU}{\raise-1pt\hbox{$\mbox{\Bbbb U}$}}

\newcommand{\z}{\raise-1pt\hbox{$\mbox{\Bbbb Z}$}}

\def\res{\mathop{\hbox{res}}\limits}

\def\beq{\begin{equation}}
\def\eeq{\end{equation}}
\def\p{\partial}

\begin{document}

\begin{titlepage}

\title{Spin generalization of the Calogero-Moser hierarchy and the matrix KP hierarchy}

\author{V.~Pashkov\thanks{Skolkovo Institute of Science and Technology, 143026 Moscow, Russian Federation; \newline
e-mail: vitaly.pashkov@skolkovotech.ru
}
\and
 A.~Zabrodin\thanks{National Research University Higher School of Economics,
20 Myasnitskaya Ulitsa, Moscow 101000, Russian Federation;
ITEP, 25 B.Cheremushkinskaya, Moscow 117218, Russian Federation;
e-mail: zabrodin@itep.ru}
}

\date{December 2017}
\maketitle

\vspace{-7cm} \centerline{ \hfill ITEP-TH-40/17}\vspace{7cm}

\begin{abstract}

We establish a correspondence between rational solutions to the matrix KP
hierarchy and the spin generalization of the Calogero-Moser system on the level 
of hierarchies. Namely, it is shown that the rational solutions to the matrix KP
hierarchy appear to be isomorphic to the spin Calogero-Moser system in a sense that
the dynamics of poles of solutions to the matrix KP hierarchy in the higher times 
is governed by the higher 
Hamiltonians of the spin Calogero-Moser integrable hierarchy with rational potential.

\end{abstract}

\end{titlepage}

\vspace{5mm}

\section{Introduction}

In the paper \cite{AMM77} it was discovered that the motion of poles of rational solutions to the
Korteweg-de Vries (KdV) and Boussinesq equations is given by dynamics of the 
many-body Calogero-Moser system of particles \cite{Calogero71,Calogero75,Moser75} 
with some additional restrictions in the configuration space.
Subsequently, the celebrated isomorphism between dynamics of poles of rational solutions
to the Kadomtsev-Petviashvili (KP) equation (which generalizes the KdV and Boussinesq equations)
and solutions to equations of motion for the rational Calogero-Moser system was established 
in \cite{Krichever78} (see also \cite{CC77}). 
Namely, in \cite{Krichever78} the general approach of 
constructing rational solutions in the variable $t_1$ to the KP equation
was proposed and it was found that the positions of the poles $x_i$ change with time
$t_2$ in the same way as the particles of the rational Calogero-Moser system.
This remarkable connection was further generalized by Krichever in \cite{Krichever80}, where
the analogous results were obtained in the case of elliptic solutions.

The further development is Shiota's work \cite{Shiota94}, where the correspondence 
between dynamics of poles of rational KP solutions and many-body integrable systems of particles was
extended to the level of {\it hierarchies}. There it was proved that the evolution of poles with respect to the
higher times $t_k$ of the infinite KP hierarchy is governed by higher Hamiltonians $H_k$ of the 
integrable Calogero-Moser system. 

In this note we generalize this result to the rational solutions of the matrix KP hierarchy.
It should be noted that singular (in general, elliptic) solutions to the matrix KP equation 
were studied in \cite{KBBT95}. It has been shown that the evolution of data of such solutions
(positions of poles and some internal degrees of freedom)
with respect to the time $t_2$ is isomorphic to the dynamics of a spin generalization of the Calogero-Moser 
system. This generalization is known as the Gibbons-Hermsen system \cite{GH84}. It is a system of ${\cal N}$
particles with coordinates $x_i$ with internal degrees of freedom 
given by $N$-dimensional column vectors ${\bf a}_i, {\bf b}_i$ which pairwise interact with each other.  
The Hamiltonian is
$$
H=\sum_{i=1}^{{\cal N}}p_i^2-\sum_{i\neq k}
\frac{({\bf b}_i^T{\bf a}_k)({\bf b}_k^T{\bf a}_i)}{(x_i-x_k)^2}
$$
(here ${\bf b}_i^T$ is the transposed row-vector) with the non-vanishing Poisson brackets
$
\{x_i, p_k\}=\delta_{ik}, \, \{a_i^{\alpha}, b_k^{\beta}\}=
\delta_{\alpha \beta}\delta_{ik}
$.
The model is known to be integrable, with the higher Hamiltonians in involution being given by
$H_k=\mbox{tr} \, L^k$, where $L$ is the Lax matrix of the model.

Here we extend this result to the level of hierarchies, i.e., we show that
the evolution of the poles and the internal degrees of freedom with respect to the
higher times $t_k$ of the matrix KP hierarchy is governed by the higher Hamiltonians $H_k$ of the 
Gibbons-Hermsen system.

The matrix extension of the KP hierarchy is closely related to the so-called
multicomponent KP hierarchy \cite{DJKM81,KL93}. In section 2, we start with a short review 
of these hierarchies. We use the bilinear formalism. The main object (the dependent variable)
is the tau-function $\tau$ which obeys an infinite number of bilinear relations encoded
by the universal bilinear identity (\ref{m1}). We also introduce the matrix Baker-Akhiezer functions 
$\Psi$, $\Psi^{\dag}$
which satisfy a system of linear equations. The compatibility conditions of this system give 
non-linear equations of the hierarchy. In section 3, we study the rational solutions of the matrix
KP hierarchy in the time $t_1$. For such solutions, the tau-function is a polynomial in $x=t_1$ with roots $x_i$,
with the Baker-Akhiezer functions having simple poles at the points $x_i$. Using the bilinear equations for 
the tau-function, we show that residues at these poles are matrices of rank 1. The internal degrees of
freedom associated with $x_i$ are expressed in terms of these residues. The dynamics of $x_i$ 
and the internal degrees of freedom is derived using the linear problems for the Baker-Akhiezer
functions.
It should be noted that
rational solutions to the multicomponent and matrix KP hierarchies and their relation to
Calogero-like systems were studied
in \cite{T11,BGK09,BZT11,BZT08,CS17} from other perspectives and points of view.

\section{The matrix KP hierarchy}

\subsection{The bilinear identity for the multicomponent KP hierarchy}

First of all, consider the multicomponent KP hierarchy \cite{DJKM81,KL93}. 
In the bilinear formalism, it can be
defined as follows (see, e.g., \cite{Teo11,TT07}). Suppose there are $N$ 
infinite sets of the independent continuous time
variables:
$$
{\bf t}=\{{\bf t}_1, {\bf t}_2, \ldots , {\bf t}_N\}, \qquad
{\bf t}_{\alpha}=\{t_{\alpha , 1}, t_{\alpha , 2}, t_{\alpha , 3}, \ldots \, \},
\qquad \alpha = 1, \ldots , N.
$$
Next, one introduces $N$ discrete variables called charges
$$
{\bf s}=\{s_1, s_2, \ldots , s_N\}, \qquad \sum_{\alpha =1}^N s_{\alpha}=0
$$
(they are integer numbers).
The $N$-component KP hierarchy is then defined by the infinite set of bilinear equations
for the tau-function $\tau ({\bf s}; {\bf t})$ that follow from the condition (the bilinear
identity)
\beq\label{m1}
\begin{array}{l}
\displaystyle{\sum_{\gamma =1}^N \epsilon_{\alpha \gamma}({\bf s})\epsilon_{\beta \gamma}({\bf s}')
\oint_{C_{\infty}}dz \, 
z^{s_{\gamma}-s_{\gamma}'+\delta_{\alpha \gamma}+\delta_{\beta \gamma}-2}
e^{\xi ({\bf t}_{\gamma}-{\bf t}_{\gamma}', \, z)}}
\\ \\
\displaystyle{\hspace{1cm}
\cdot \tau \left ({\bf s}+{\bf e}_{\alpha}-{\bf e}_{\gamma}; {\bf t}-[z^{-1}]_{\gamma}\right )
\tau \left ({\bf s}'+{\bf e}_{\gamma}-{\bf e}_{\beta}; {\bf t}'+[z^{-1}]_{\gamma}\right )=0,
\qquad \alpha, \beta =1, \ldots , N,}
\end{array}
\eeq
valid for any ${\bf s}$, ${\bf s}'$, ${\bf t}$, ${\bf t}'$. The notation is as follows:
${\bf e}_{\alpha}$ is the vector with 1 on the $\alpha$th place and with all other entries equal to zero,
$$
\epsilon_{\alpha \gamma}({\bf s})=\left \{
\begin{array}{ll}
\;\; (-1)^{s_{\alpha +1}+\ldots +s_{\gamma}} &\quad \mbox{if $\alpha <\gamma$}
\\
\quad 1 &\quad \mbox{if $\alpha =\gamma$}
\\
-(-1)^{s_{\gamma +1}+\ldots +s_{\alpha}} &\quad \mbox{if $\alpha >\gamma$}
\end{array}\right.
$$
and
$$
\xi ({\bf t}_{\gamma}, z)=\sum_{k\geq 1}t_{\gamma , k}z^k,
$$
$$
\left ({\bf t}\pm [z^{-1}]_{\gamma}\right )_{\alpha k}=t_{\alpha , k}\pm
\delta_{\alpha \gamma} \frac{z^{-k}}{k}.
$$
The integration contour $C_{\infty}$ around $\infty$ is such that all singularities 
coming from the power of $z$ and the exponential function
$e^{\xi ({\bf t}_{\gamma}-{\bf t}_{\gamma}', \, z)}$ are inside it and all singularities 
coming from the $\tau$-factors are outside it. We remark that the sign factors 
$\epsilon_{\alpha \beta }({\bf s})$ satisfy the identities
\beq\label{e1}
\epsilon_{\beta \alpha}({\bf s})=-\epsilon_{\alpha \beta }({\bf s}), \qquad
\epsilon_{\alpha \beta }(-{\bf s})=\epsilon_{\alpha \beta }({\bf s}), \qquad
\epsilon_{\alpha \gamma }({\bf s}+{\bf e}_{\alpha}-{\bf e}_{\beta})=
\epsilon_{\beta \gamma}({\bf s})\epsilon_{\beta \alpha}({\bf s})
\eeq
for any distinct $\alpha , \beta , \gamma$.

\subsection{The Hirota equations for the multicomponent KP hierarchy}

Choosing ${\bf s}'$ and ${\bf t}'$ in (\ref{m1}) in a specific way, one can obtain,
after calculating the integral with the help of residues, a number of differential and
difference Hirota bilinear equations for the tau-function (called Fay identities in \cite{Teo11}).
The full list of such equations is given in \cite{Teo11}. Here we give only the equations that 
are used in what follows. 

For any distinct $\alpha , \beta , \kappa$ it holds
\beq\label{h1}
\begin{array}{c}
\p_{t_{\kappa , 1}}\tau ({\bf s}, {\bf t})\cdot 
\tau ({\bf s}+{\bf e}_{\alpha}-{\bf e}_{\beta}; {\bf t}) -
\p_{t_{\kappa , 1}}\tau ({\bf s}+{\bf e}_{\alpha}-{\bf e}_{\beta}; {\bf t})\cdot
\tau ({\bf s}, {\bf t})
\\ \\
\displaystyle{=-\frac{\epsilon_{\alpha \kappa}({\bf s})
\epsilon_{\beta \kappa}({\bf s})}{\epsilon_{\beta \alpha}({\bf s})}\,
\tau ({\bf s}+{\bf e}_{\alpha}-{\bf e}_{\kappa}; {\bf t})\cdot
\tau ({\bf s}+{\bf e}_{\kappa}-{\bf e}_{\beta}; {\bf t})}.
\end{array}
\eeq
For any distinct $\alpha , \beta$ it holds
\beq\label{h2}
\begin{array}{c}
\p_{t_{\alpha , 1}}\tau ({\bf s}+{\bf e}_{\alpha}-{\bf e}_{\beta}; {\bf t})\cdot
\tau \left ({\bf s}; {\bf t}-[z^{-1}]_{\alpha}\right )-
\p_{t_{\alpha , 1}}\tau \left ({\bf s}; {\bf t}-[z^{-1}]_{\alpha}\right )\cdot
\tau ({\bf s}+{\bf e}_{\alpha}-{\bf e}_{\beta}; {\bf t})
\\ \\
\displaystyle{=z\tau \left ({\bf s}; {\bf t}-[z^{-1}]_{\alpha}\right )\cdot
\tau ({\bf s}+{\bf e}_{\alpha}-{\bf e}_{\beta}; {\bf t})-z
\tau ({\bf s}; {\bf t})\cdot 
\tau \left ({\bf s}+{\bf e}_{\alpha}-{\bf e}_{\beta}; {\bf t}-[z^{-1}]_{\alpha}\right )}.
\end{array}
\eeq
Taking ${\bf s}'={\bf s}+{\bf e}_{\kappa}-{\bf e}_{\lambda}$, ${\bf t}'={\bf t}$ in the
bilinear identity, one can see that for any distinct $\alpha , \beta , \kappa , \lambda$ it holds
\beq\label{h3}
\begin{array}{c}
\epsilon_{\beta \alpha}({\bf s}+{\bf e}_{\kappa}-{\bf e}_{\lambda})
\tau ({\bf s}; {\bf t})\cdot
\tau ({\bf s}+{\bf e}_{\kappa}-{\bf e}_{\lambda}+{\bf e}_{\alpha}-{\bf e}_{\beta}; {\bf t})
\\ \\
+\, \epsilon_{\alpha \beta}({\bf s})
\tau ({\bf s}+{\bf e}_{\alpha}-{\bf e}_{\beta}; {\bf t})\cdot
\tau ({\bf s}+{\bf e}_{\kappa}-{\bf e}_{\lambda}; {\bf t})
\\ \\
+\, \epsilon_{\alpha \lambda}({\bf s})\epsilon_{\beta \lambda}({\bf s}+{\bf e}_{\kappa}-{\bf e}_{\lambda})
\tau ({\bf s}+{\bf e}_{\alpha}-{\bf e}_{\lambda}; {\bf t})\cdot
\tau ({\bf s}+{\bf e}_{\kappa}-{\bf e}_{\beta}; {\bf t})=0.
\end{array}
\eeq

\subsection{The Baker-Akhiezer functions}

The Baker-Akhiezer function $\Psi ({\bf s}, {\bf t};z)$ and its adjoint
$\Psi ^{*}({\bf s}, {\bf t};z)$ are $N\! \times \! N$ matrices with components
defined by the following formulae:
\beq\label{m2}
\begin{array}{l}
\displaystyle{\Psi_{\alpha \beta}({\bf s}, {\bf t};z)=
\epsilon_{\alpha \beta}({\bf s})\,
\frac{\tau \left (
{\bf s}\! +\! {\bf e}_{\alpha}\! -\! {\bf e}_{\beta}; 
{\bf t}-[z^{-1}]_{\beta}\right )}{\tau ({\bf s}; {\bf t})}\,
z^{s_{\beta}+\delta_{\alpha \beta}-1}e^{\xi ({\bf t}_{\beta}, z)},
}
\\ \\
\displaystyle{\Psi_{\alpha \beta}^*({\bf s}, {\bf t};z)=
\epsilon_{\alpha \beta}({\bf s})\,
\frac{\tau \left (
{\bf s}\! -\! {\bf e}_{\alpha}\! +\! {\bf e}_{\beta}; 
{\bf t}+[z^{-1}]_{\beta}\right )}{\tau ({\bf s}; {\bf t})}\,
z^{-s_{\beta}+\delta_{\alpha \beta}-1}e^{-\xi ({\bf t}_{\beta}, z)}.
}
\end{array}
\eeq
In terms of the Baker-Akhiezer functions, the bilinear identity (\ref{m1}) can be written
as
\beq\label{m3}
\oint_{C_{\infty}}\! dz \, \Psi ({\bf s}, {\bf t};z)\Psi^{\dag} ({\bf s}', {\bf t}';z)=0,
\qquad
\Psi^{\dag}_{\alpha \beta} = \Psi^{*}_{\beta \alpha}
\eeq
(here and below $\Psi^{\dag}$ does not mean the Hermitian conjugation).
Around $z=\infty$, the Baker-Akhiezer functions can be represented in the form of the series
\beq\label{m4}
\Psi_{\alpha \beta}({\bf s}, {\bf t};z)=\left (\delta_{\alpha \beta}+
\sum_{k\geq 1}\frac{w^{(k)}_{\alpha \beta}({\bf s}, 
{\bf t})}{z^k}\right )z^{s_{\beta}}e^{\xi ({\bf t}_{\beta}, z)},
\eeq
\beq\label{m4a}
\Psi^{*}_{\alpha \beta}({\bf s}, {\bf t};z)=\left (\delta_{\alpha \beta}+
\sum_{k\geq 1}\frac{w^{*(k)}_{\alpha \beta}({\bf s}, 
{\bf t})}{z^k}\right )z^{-s_{\beta}}e^{-\xi ({\bf t}_{\beta}, z)}.
\eeq
It follows from the bilinear identity in the form (\ref{m3}) taken at ${\bf s}'={\bf s}$,
${\bf t}'={\bf t}$ that 
\beq\label{m4b}
w^{*(1)}_{\alpha \beta}({\bf s}, {\bf t})=-
w^{(1)}_{\alpha \beta}({\bf s}, {\bf t}).
\eeq

The multicomponent KP hierarchy can be understood as an infinite set of 
evolution equations in the times ${\bf t}$ for matrix functions of a variable $x$. 
For example, one can consider the coefficients $w^{(k)}$ of the Baker-Akhiezer function 
as such matrix functions, the evolution being $w^{(k)}(x)\to w^{(k)}(x,{\bf t})$.
In what follows we will denote $\tau (x, {\bf t})$, $w^{(k)}(x,{\bf t})$ simply as 
$\tau ({\bf t})$, $w^{(k)}({\bf t})$. 
Let us introduce the (matrix pseudo-differential) wave operator 
$$
W=I+\sum_{k\geq 1}
w^{(k)}({\bf t})\p_x^{-k},
$$
where $I$ is the unity $N\! \times \! N$ matrix and $w^{(k)}({\bf t})$ are the
same matrix functions as in (\ref{m4}). Writing this in 
matrix elements, we have
\beq\label{m113}
W_{\alpha \beta} = \delta_{\alpha \beta}+\sum_{k\geq 1}
w^{(k)}_{\alpha \beta}({\bf t})\p_x^{-k}.
\eeq
The Baker-Akhiezer function can be written as a result of action of the wave operator 
to the exponential function:
$$
\Psi ({\bf t}; z)=W\exp \Bigl (xzI+\sum_{\alpha =1}^N E_{\alpha}\xi ({\bf t}_{\alpha}, z)\Bigr ),
$$
where $E_{\alpha}$ is the $N\! \times \! N$ matrix with $1$ on the $(\alpha , \alpha )$ component
and zero elsewhere. 
The adjoint Baker-Akhiezer function can be written as
$$
\Psi ^{\dag} ({\bf t}; z)=\exp \Bigl (-xzI-\sum_{\alpha =1}^N E_{\alpha}\xi ({\bf t}_{\alpha}, z)\Bigr )
W^{-1}.
$$
Here it is assumed that the operators $\p_x$ 
entering $W^{-1}$ act to the left (i.e., we define $f\p_x =-\p_x f$).

As is proved in \cite{Teo11}, the Baker-Akhiezer function and its adjoint 
satisfy the linear
equations
\beq\label{m13c}
\begin{array}{l}
\,\,\,\,\, \p_{t_{\alpha , m}}\Psi ({\bf t}; z)=B_{\alpha m} \Psi ({\bf t}; z), 
\\ \\
-\p_{t_{\alpha , m}}\Psi^{\dag} ({\bf t}; z)=\Psi^{\dag} ({\bf t}; z) B_{\alpha m} , 
\end{array}
\eeq
where $B_{\alpha m}$ is the differential operator
$$
B_{\alpha m}= \Bigl (W E_{\alpha}\p_x^m W^{-1}\Bigr )_+.
$$
Here $(\ldots )_+$ denotes the differential part of a pseudo-differential operator, i.e.
the sum of terms with $\p_x^k$, where $k\geq 0$. In particular, 
\beq\label{m13d}
\sum_{\alpha =1}^{N}\p_{t_{\alpha , 1}}\Psi ({\bf t}; z)=\p_x \Psi ({\bf t}; z),
\eeq
so the vector field $\p_x$ can be identified with the vector field
$\sum_{\alpha }\p_{t_{\alpha , 1}}$.

\subsection{The matrix KP hierarchy and linear problems for the Ba\-ker-\-Akhi\-ez\-er functions}

Let us proceed to the specification to the matrix KP hierarchy. 
This hierarchy
results from the multicomponent KP one 
after a restriction of the times and the charge
variables. For each set of the times ${\bf t}_{\alpha}$ we fix the ``initial values''
${\bf t}_{\alpha}^{(0)}$ and suppose that the times change in the following manner:
$$
t_{\alpha , m}=t_{\alpha , m}^{(0)}+ t_m \qquad \mbox{for each $\alpha$ and $m$}.
$$
In other words, for any fixed $m$, the time evolution with respect to each $t_{\alpha , m}$
is the same and is defined by $t_m$ only. The corresponding vector fields are related as
$\p_{t_m}=\sum_{\alpha =1}^N \p_{t_{\alpha , m}}$.
The charge variables are supposed to be fixed.
It is convenient to put ${\bf s}=0$.
In what follows we omit them in the notation for the tau-function and the Baker-Akhiezer functions and 
put ${\bf s}={\bf s}'=0$ in the bilinear identity. Accordingly, the bilinear identity for the 
matrix KP hierarchy acquires the form
\beq\label{m5}
\sum_{\gamma =1}^N \epsilon_{\alpha \gamma}\epsilon_{\beta \gamma}
\oint_{C_{\infty}}dz \, 
z^{\delta_{\alpha \gamma}+\delta_{\beta \gamma}-2}
e^{\xi ({\bf t}_{\gamma}-{\bf t}_{\gamma}', \, z)}
\tau _{\alpha \gamma} \left ({\bf t}-[z^{-1}]_{\gamma}\right )
\tau _{\gamma \beta}\left ({\bf t}'+[z^{-1}]_{\gamma}\right )=0,
\eeq 
where $\epsilon_{\alpha \gamma}=1$ if $\alpha \leq \gamma$, $\epsilon_{\alpha \gamma}=-1$
if $\alpha >\gamma$ and 
\beq\label{m6}
\tau _{\alpha \beta}({\bf t})=\tau ({\bf e}_{\alpha}-{\bf e}_{\beta}; {\bf t}).
\eeq
The Baker-Akhiezer function and its adjoint  have the expansions
\beq\label{m7}
\begin{array}{l}
\Psi_{\alpha \beta}({\bf t};z)=\left (\delta_{\alpha \beta}+
w_{\alpha \beta}^{(1)}({\bf t})z^{-1}+O(z^{-2})\right )
e^{xz+\xi ({\bf t}, z)}
\\ \\
\Psi^{*}_{\alpha \beta}({\bf t};z)=\left (\delta_{\alpha \beta}-
w_{\alpha \beta}^{(1)}({\bf t})z^{-1}+O(z^{-2})\right )
e^{-xz-\xi ({\bf t}, z)},
\end{array}
\eeq
where $\displaystyle{\xi ({\bf t}, z)=\sum_{k\geq 1}t_kz^k}$.
It is easy to see from (\ref{m2}) that
\beq\label{m8}
w^{(1)}_{\alpha \beta}({\bf t})=\left \{
\begin{array}{l}
\displaystyle{\epsilon_{\alpha \beta}\, \frac{\tau_{\alpha \beta}({\bf t})}{\tau ({\bf t})}}\qquad
\,\, \mbox{if $\alpha \neq \beta$}
\\ \\
\displaystyle{-\, \frac{\p_{t_{\alpha , 1}}\tau ({\bf t})}{\tau ({\bf t})}} \qquad 
\mbox{if $\alpha = \beta$.}
\end{array}\right.
\eeq

Let us derive a useful corollary of the bilinear identity which will be used for analysis
of the rational solutions. In order to obtain it, we differentiate the bilinear
identity with respect to $t_m$ and put ${\bf t}'={\bf t}$ after this. It is not difficult
to see that the result is
\beq\label{m9}
\frac{1}{2\pi i}
\sum_{\gamma =1}^N \oint_{C_{\infty}}dz \, z^m \Psi_{\alpha \gamma}({\bf t}; z)
\Psi^{*}_{\beta \gamma}({\bf t}; z)=-\epsilon_{\alpha \beta}\,\p_{t_m}\! \!\left (
\frac{\tau_{\alpha \beta}({\bf t})}{\tau ({\bf t})}\right )
\eeq
for $\alpha \neq \beta$ and
\beq\label{m10}
\frac{1}{2\pi i}
\sum_{\gamma =1}^N \oint_{C_{\infty}}dz \, z^m \Psi_{\alpha \gamma}({\bf t}; z)
\Psi^{*}_{\alpha \gamma}({\bf t}; z)=\p_{t_m}\p_{t_{\alpha , 1}}\log \tau ({\bf t}).
\eeq
Comparing with (\ref{m8}), we conclude that
\beq\label{m11}
\frac{1}{2\pi i}
\sum_{\gamma =1}^N \oint_{C_{\infty}}dz \, z^m \Psi_{\alpha \gamma}({\bf t}; z)
\Psi^{*}_{\beta \gamma}({\bf t}; z)=-\p_{t_m}w_{\alpha \beta}^{(1)}({\bf t})
\eeq
for any $\alpha , \beta$.
Note also that summing (\ref{m10}) over $\alpha$ from $1$ to $N$, we get
\beq\label{m12}
\frac{1}{2\pi i}
\sum_{\alpha, \beta =1}^N \oint_{C_{\infty}}\!\! dz \, z^m \Psi_{\alpha \beta}({\bf t}; z)
\Psi^{*}_{\alpha \beta}({\bf t}; z)=
\frac{1}{2\pi i}
\oint_{C_{\infty}}\!\! dz \, z^m \mbox{tr}\, \Psi ({\bf t}; z)\Psi^{\dag}({\bf t}; z)
=\p_{t_m}\p_{t_{1}}\log \tau ({\bf t}).
\eeq

Recalling (\ref{m13d}), we can identify
$$
\p_x = \p_{t_1}=\sum_{\alpha =1}^N \p_{t_{\alpha , 1}}.
$$
As it follows from (\ref{m13c}), 
the Baker-Akhiezer function and its adjoint 
satisfy the linear
equations
\beq\label{m13a}
\begin{array}{l}
\,\,\,\,\, \p_{t_m}\Psi ({\bf t}; z)=B_m \Psi ({\bf t}; z), \qquad \,\,\,\, m\geq 1,
\\ \\
-\p_{t_m}\Psi^{\dag} ({\bf t}; z)=\Psi^{\dag} ({\bf t}; z) B_m , \qquad m\geq 1,
\end{array}
\eeq
where $B_m$ is the differential operator
$$
B_m= \Bigl (W \p_x^m W^{-1}\Bigr )_+.
$$
At $m=1$ we have $\p_{t_1}\Psi =\p_x \Psi$, so the evolution in $t_1$ is simply a shift of
the variable $x$:
\beq\label{m16}
w^{(k)}(x, t_1, t_2, \ldots )=w^{(k)}(x+t_1, t_2, \ldots ).
\eeq
At $m=2$ we have the linear problems
\beq\label{m14}
\p_{t_2}\Psi = \p_x^2\Psi +V({\bf t})\Psi ,  
\eeq
\beq\label{m14a}
-\p_{t_2}\Psi^{\dag} = \p_x^2\Psi^{\dag} +\Psi^{\dag}V({\bf t})
\eeq
which have the form of the non-stationary matrix Schrodinger equations with the potential
\beq\label{m15}
V({\bf t})=-2\p_x w^{(1)}({\bf t}).
\eeq

\section{Rational solutions to the matrix KP hierarchy}

In this section we study solutions to the matrix KP hierarchy which are rational functions
of the variable $x$ (and, therefore, $t_1$). First of all 
we find the form of the Baker-Akhiezer functions for the 
rational solutions.

\subsection{Baker-Akhiezer functions for rational solutions}

For the rational solutions, the tau-function should be a polynomial in $x$ (possibly 
multiplied by an exponential function):
\beq\label{r1}
\tau ({\bf t})= Ce^{Ax}\prod_{i=1}^{{\cal N}} (x-x_i({\bf t})).
\eeq
Here ${\cal N}$ is the number of roots $x_i$ of the polynomial and the roots depend 
on the times ${\bf t}$. We assume that all the roots are distinct. Let us use the notation
$$
\tau '(x_i)=\lim_{x\to x_i}\frac{\tau ({\bf t})}{x-x_i}=Ce^{Ax_i}\prod_{j\neq i}(x_i-x_j),
$$
$$
\tau_{\alpha \beta}(x_i)=\tau_{\alpha \beta}({\bf t})\Bigr |_{x=x_i}, \qquad
\p_{t_{\alpha , 1}}\tau (x_i)=\p_{t_{\alpha , 1}}\tau ({\bf t})\Bigr | _{x=x_i}.
$$
It is clear from (\ref{m2}) that the Baker-Akhiezer functions $\Psi$, $\Psi^*$, as functions of $x$, 
have simple poles at $x=x_i$.
From (\ref{m8}) we see that the residue of $w_{\alpha \beta}^{(1)}$ (as a function of $x$)
at the pole $x_i$ is given by
\beq\label{r2}
\res_{x=x_i}\! w_{\alpha \beta}^{(1)}=\left \{
\begin{array}{l}
\displaystyle{ \epsilon_{\alpha \beta}\frac{\tau_{\alpha \beta}(x_i)}{\tau '(x_i)}\qquad
\,\,\, \mbox{if $\alpha \neq \beta$},}
\\ \\
\displaystyle{ 
-\,\frac{\p_{t_{\alpha , 1}}\tau (x_i)}{\tau '(x_i)} \qquad
\mbox{if $\alpha = \beta$}.}
\end{array}\right.
\eeq
We are going to show, using the Hirota equations (\ref{h1})--(\ref{h3}), that the 
dependence on $\alpha$ and $\beta$ in $\res_{x=x_i}w_{\alpha \beta}^{(1)}$ actually factorizes, i.e.,
\beq\label{r3}
\res_{x=x_i}\! w_{\alpha \beta}^{(1)}=-a_i^{\alpha}b_{i}^{\beta} \qquad
\mbox{or} \qquad
\res_{x=x_i}\! w^{(1)}=-{\bf a}_i {\bf b}_i^{T}
\eeq
for some column vectors ${\bf a}_i =(a_i^1, a_i^2, \ldots , a_i^N)^T$,
${\bf b}_i =(b_i^1, b_i^2, \ldots , b_i^N)^T$ ($T$ means transposition), so the matrix
$\res_{x=x_i}\! w^{(1)}$ is of rank $1$.
Note that in \cite{KBBT95} the form (\ref{r3}) was derived from some
algebro-geometric considerations using analytic properties of the Baker-Akhiezer function
on the algebraic curve. 

Setting in (\ref{h3}) ${\bf s}=0$ and taking it at $x=x_i$ (so that the first term vanishes),
we arrive at the relation
$$
\epsilon_{\alpha \beta}\tau_{\alpha \beta}(x_i)=
\epsilon_{\alpha \lambda}\epsilon_{\kappa \beta}\epsilon_{\kappa \lambda}\,
\frac{\tau_{\alpha \lambda}(x_i)\tau _{\kappa \beta}(x_i)}{\tau_{\kappa \lambda}(x_i)}
$$
for distinct $\alpha , \beta , \kappa , \lambda$,
where (\ref{e1}) was used for the transformation of $\epsilon$-factors.
Consider now (\ref{h1}), put ${\bf s}=0$ there, change $\alpha \to \kappa$,
$\kappa \to \lambda$ and substitute
$x=x_i$ (so that the second term in the left hand side vanishes). We get
$$
\epsilon_{\lambda \beta}\tau_{\lambda \beta}(x_i)=-\epsilon_{\kappa \beta}
\epsilon_{\kappa \lambda}\, \frac{\p_{t_{\lambda , 1}}\tau (x_i) 
\tau_{\kappa \beta}(x_i)}{\tau_{\kappa \lambda}(x_i)}.
$$
Similarly, changing in (\ref{h1}) $\beta \to \lambda$ and putting $x=x_i$
(the second term in the left hand side vanishes), we get
$$
\epsilon_{\alpha \kappa}\tau_{\alpha \kappa}(x_i)=-\epsilon_{\alpha \lambda}
\epsilon_{\kappa \lambda}\, \frac{\p_{t_{\kappa , 1}}\tau (x_i) 
\tau_{\alpha \lambda}(x_i)}{\tau_{\kappa \lambda}(x_i)}.
$$
Altogether, these formulae mean that
\beq\label{r4}
\epsilon_{\alpha \beta}\tau_{\alpha \beta}(x_i)=\frac{A_{\alpha}B_{\beta}}{\epsilon_{\kappa \lambda}
\tau_{\kappa \lambda}(x_i)}, \qquad \alpha \neq \beta ,
\eeq
i.e., the factorization holds for $\alpha \neq \beta$ with
$$
A_{\alpha}=\left \{ \begin{array}{l}
\epsilon_{\alpha \lambda}\tau_{\alpha \lambda}(x_i), \qquad \,\, \alpha \neq \lambda
\\ \\
-\p_{t_{\lambda , 1}}\tau (x_i), \qquad \alpha =\lambda \, ,
\end{array}\right.
$$
$$
B_{\beta}=\left \{ \begin{array}{l}
\epsilon_{\kappa \beta}\tau_{\kappa \beta}(x_i), \qquad \,\, \beta \neq \kappa
\\ \\
-\p_{t_{\kappa , 1}}\tau (x_i), \qquad \beta =\kappa \, .
\end{array}\right.
$$
Moreover, at $\alpha =\beta$ we use (\ref{h1}) with the changes
$\kappa \to \alpha$, $\alpha \to \kappa$, $\beta \to \lambda$. At 
$x=x_i$ (the second term in the left hand side vanishes) we get
$$
-\p_{t_{\alpha , 1}}\tau (x_i)=\epsilon_{\alpha \lambda}
\epsilon_{\kappa \alpha}\epsilon_{\kappa \lambda}\,  
\frac{\tau_{\alpha \lambda}(x_i)\tau_{\kappa \alpha}(x_i)}{\tau_{\kappa \lambda}(x_i)},
$$
which means, together with (\ref{r4}), that 
$$
\res_{x=x_i}\! w_{\alpha \beta}^{(1)}=\epsilon_{\kappa \lambda}\,
\frac{A_{\alpha}\, B_{\beta}}{\tau_{\kappa \lambda}(x_i)\tau '(x_i)} \qquad \mbox{for any
$\alpha , \beta$},
$$
so the representation (\ref{r3}) is valid.

Now we turn to the residues of the Baker-Akhiezer functions. From (\ref{m2}) at
${\bf s}=0$ and $t^{(0)}_{\alpha , m}=0$ we have:
\beq\label{r5}
\res_{x=x_i}\! \Psi_{\alpha \beta}=e^{xz+\xi ({\bf t}, z)}z^{\delta_{\alpha \beta}-1}
\epsilon_{\alpha \beta}\, \frac{\tau_{\alpha \beta}(x_i ; {\bf t}-[z^{-1}]_{\beta})}{\tau ' (x_i)},
\eeq
\beq\label{r6}
\res_{x=x_i}\! \Psi_{\alpha \beta}^{\dag}=-e^{-xz-\xi ({\bf t}, z)}z^{\delta_{\alpha \beta}-1}
\epsilon_{\alpha \beta}\, \frac{\tau_{\alpha \beta}(x_i ; {\bf t}+[z^{-1}]_{\alpha})}{\tau ' (x_i)}
\eeq
for any $\alpha , \beta$.
In order to transform these expressions, we use the Hirota equation (\ref{h2}). 
First, we set ${\bf s}={\bf e}_{\beta}-{\bf e}_{\alpha}$, change $\alpha \leftrightarrow \beta$ and
substitute $x=x_i$, so that the second term in the left hand side and the first term in the 
right hand side vanish. The result is
$$
z^{-1}\tau_{\alpha \beta}(x_i, {\bf t}-[z^{-1}]_{\beta})=-\tau_{\alpha \beta}(x_i, {\bf t})\,
\frac{\tau (x_i, {\bf t}-[z^{-1}]_{\beta})}{\p_{t_{\beta , 1}}\tau (x_i , {\bf t})}.
$$
Similarly, changing ${\bf t}\to {\bf t}+[z^{-1}]_{\alpha}$ in (\ref{h2}) and putting $x=x_i$, 
we get
$$
z^{-1}\tau_{\alpha \beta}(x_i, {\bf t}+[z^{-1}]_{\alpha})=\tau_{\alpha \beta}(x_i, {\bf t})\,
\frac{\tau (x_i, {\bf t}+[z^{-1}]_{\alpha})}{\p_{t_{\alpha , 1}}\tau (x_i , {\bf t})}.
$$
Using these formulae, it is easy to see that equations (\ref{r5}), (\ref{r6})
can be written in the form
$$
\res_{x=x_i}\! \Psi_{\alpha \beta}=-\left (\res_{x=x_i}\! w^{(1)}_{\alpha \beta}\right )
e^{xz+\xi ({\bf t}, z)}\frac{\tau (x_i ; {\bf t}-[z^{-1}]_{\beta})}{\p_{t_{\beta , 1}}
\tau (x_i; {\bf t})},
$$
$$
\res_{x=x_i}\! \Psi_{\alpha \beta}^{\dag}=-\left (\res_{x=x_i}\! w^{(1)}_{\alpha \beta}\right )
e^{-xz-\xi ({\bf t}, z)}\frac{\tau (x_i ; {\bf t}+[z^{-1}]_{\alpha})}{\p_{t_{\alpha , 1}}
\tau (x_i; {\bf t})}
$$
for any $\alpha , \beta$. Therefore, plugging here (\ref{r3}), we conclude that 
$\res_{x=x_i}\! \Psi$, $\res_{x=x_i}\! \Psi ^{\dag}$ are matrices of rank $1$:
\beq\label{r7}
\res_{x=x_i}\! \Psi_{\alpha \beta}=e^{xz+\xi ({\bf t}, z)}a_i^{\alpha}c_{i}^{\beta},
\qquad
\res_{x=x_i}\! \Psi_{\alpha \beta}^{\dag}=e^{-xz-\xi ({\bf t}, z)}c_i^{*\alpha}b_{i}^{\beta},
\eeq
where $c_i^{\alpha}$, $c_i^{*\alpha}$ are components of some 
vectors ${\bf c}_i=(c_i^1, \ldots , c_i^N)^T$, 
${\bf c}^{*}_i=(c^{*1}_i, \ldots , c^{*N}_i)^T$.

Summing up, we have the following representation of the Baker-Akhiezer functions:
\beq\label{r8}
\Psi ({\bf t}; z)=e^{xz+\xi ({\bf t}, z)} \left ( I+\sum_{i=1}^{{\cal N}}
\frac{{\bf a}_i{\bf c}_i^T}{x-x_i({\bf t})}\right ),
\eeq
\beq\label{r9}
\Psi ^{\dag}({\bf t}; z)=e^{-xz-\xi ({\bf t}, z)} \left ( I+\sum_{i=1}^{{\cal N}}
\frac{{\bf c}^{*}_i{\bf b}_i^T}{x-x_i({\bf t})}\right ),
\eeq
or, in components,
\beq\label{r10}
\Psi _{\alpha \beta} ({\bf t}; z)=e^{xz+\xi ({\bf t}, z)} \left ( \delta_{\alpha \beta}
+\sum_{i=1}^{{\cal N}}
\frac{a_i^{\alpha}c_i^{\beta}}{x-x_i({\bf t})}\right ),
\eeq
\beq\label{r11}
\Psi ^{\dag} _{\alpha \beta} ({\bf t}; z)=e^{-xz-\xi ({\bf t}, z)} \left ( \delta_{\alpha \beta}
+\sum_{i=1}^{{\cal N}}
\frac{c_i^{*\alpha}b_i^{\beta}}{x-x_i({\bf t})}\right ).
\eeq
Here the vectors ${\bf a}_i$, ${\bf b}_i$ depend on the times $t_k$ with $k\geq 2$ while the
vectors ${\bf c}_i$, ${\bf c}^{*}_i$ depend on the same set of times and on $z$.
For the matrices $w^{(1)}$ and $V=-2\p_x w^{(1)}$ we have
\beq\label{r12}
w^{(1)}=-\sum_{i=1}^{{\cal N}} \frac{{\bf a}_i{\bf b}_i^T}{x-x_i({\bf t})},
\qquad
V({\bf t})=-2\sum_{i=1}^{{\cal N}}\frac{{\bf a}_i{\bf b}_i^T}{(x-x_i({\bf t}))^2},
\eeq
or, in components,
\beq\label{r13}
w^{(1)}_{\alpha \beta}=-\sum_{i=1}^{{\cal N}} \frac{a_i^{\alpha}b^{\beta}_i}{x-x_i({\bf t})},
\qquad
V_{\alpha \beta}({\bf t})=-2\sum_{i=1}^{{\cal N}} \frac{a_i^{\alpha}b^{\beta}_i}{(x-x_i({\bf t}))^2}.
\eeq

\subsection{Equations of motion with respect to $t_2$}

According to the Krichever approach \cite{Krichever78}, the strategy is to substitute
the pole ansatz for the Baker-Akhiezer functions  (\ref{r10}), (\ref{r11})
into the linear problems (\ref{m14}), (\ref{m14a}):
$$
\p_{t_2}\Psi_{\alpha \beta}=\p_{x}^2 \Psi_{\alpha \beta}-2
\sum_{i=1}^{{\cal N}}\sum_{\gamma} \frac{a_i^{\alpha}b^{\gamma}_i}{(x-x_i)^2}\, \Psi_{\gamma \beta},
$$
$$
-\p_{t_2}\Psi_{\alpha \beta}^{\dag}=\p_{x}^2 \Psi_{\alpha \beta}^{\dag}-2\sum_{\gamma}
\Psi^{\dag}_{\alpha \gamma} \sum_{i=1}^{{\cal N}}\frac{a_i^{\gamma}b^{\beta}_i}{(x-x_i)^2}.
$$
We have:
$$
\p_{t_2}\Psi_{\alpha \beta}=z^2\Psi_{\alpha \beta}+e^{xz+\xi ({\bf t}, z)}
\sum_{i=1}^{{\cal N}} \left (
\frac{\p_{t_2}(a_{i}^{\alpha}c_i^{\beta})}{x-x_i}+
\frac{a_{i}^{\alpha}c_i^{\beta}\dot x_i}{(x-x_i)^2}\right ),
$$
where $\dot x_k =\p_{t_2}x_k$ and
$$
\p_x^2 \Psi_{\alpha \beta}=z^2\Psi_{\alpha \beta}-2ze^{xz+\xi ({\bf t}, z)}
\sum_{i=1}^{{\cal N}} \frac{a_{i}^{\alpha}c_i^{\beta}}{(x-x_i)^2}
+2e^{xz+\xi ({\bf t}, z)}\sum_{i=1}^{{\cal N}} 
\frac{a_{i}^{\alpha}c_i^{\beta}}{(x-x_i)^3},
$$
$$
\left ( \sum_{i=1}^{{\cal N}}\frac{a_{i}^{\alpha}b_i^{\gamma}}{(x-x_i)^2}\right )
\left ( \delta_{\gamma \beta}+\sum_{k=1}^{{\cal N}}
\frac{a_{k}^{\gamma}c_k^{\beta})}{x-x_k}\right )=
\sum_{i=1}^{{\cal N}}\left (\frac{a_{i}^{\alpha}b_i^{\beta}}{(x-x_i)^2}+
\frac{a_i^{\alpha}b_i^{\gamma}a_i^{\gamma}c_i^{\beta}}{(x-x_i)^3}\right )
$$
$$
+\, \sum_{i\neq k}\frac{a_k^{\alpha}b_k^{\gamma}a_i^{\gamma}c_i^{\beta}-
a_i^{\alpha}b_i^{\gamma}a_k^{\gamma}c_k^{\beta}}{(x_i-x_k)^2(x-x_i)}+
+\sum_{i\neq k}\frac{a_i^{\alpha}b_i^{\gamma}a_k^{\gamma}c_k^{\beta}}{(x_i-x_k)
(x-x_i)^2},
$$
where summation over repeated index $\gamma$ is implied. Substituting these expressions
into the linear problem and equating coefficients at the poles at $x=x_i$ of different orders,
we get the following conditions:
\begin{itemize}
\item
At $\frac{1}{(x-x_i)^3}$: \phantom{a} $b_i^{\gamma}a_i^{\gamma}=1$ or
${\bf b}_i^T {\bf a}_i =1$;
\item
At $\frac{1}{(x-x_i)^2}$: \phantom{a} $\displaystyle{
a_i^{\alpha}c_i^{\beta}\dot x_i = -2za_i^{\alpha}c_i^{\beta}-2a_i^{\alpha}b_i^{\beta}
-2\sum_{k\neq i}\frac{a_i^{\alpha}b_i^{\gamma}a_k^{\gamma}c_k^{\beta}}{x_i-x_k}}$;
\item
At $\frac{1}{x-x_i}$: \phantom{a} $\displaystyle{\p_{t_2}(a_i^{\alpha}c_i^{\beta})=
-2\sum_{k\neq i}\frac{a_k^{\alpha}b_k^{\gamma}a_i^{\gamma}c_i^{\beta}-
a_i^{\alpha}b_i^{\gamma}a_k^{\gamma}c_k^{\beta}}{(x_i-x_k)^2}}$.
\end{itemize}
The conditions coming from the second order poles can be written in the matrix form:
\beq\label{d1}
\sum_{k=1}^{{\cal N}}(zI-L)_{ik}c_k^{\alpha}=-b_i^{\alpha}, \qquad
L_{ik}=-\frac{\dot x_i}{2}\, \delta_{ik}-(1-\delta_{ik})\, \frac{{\bf b}_i^T {\bf a}_k}{x_i-x_k},
\eeq
where $I$ is the ${\cal N}\! \times \! {\cal N}$ unity matrix. As for the conditions 
at the first order poles, we write
$\p_{t_2}(a_i^{\alpha}c_i^{\beta})=\dot a_i^{\alpha}c_i^{\beta}+a_i^{\alpha}\dot c_i^{\beta}$,
and equate the two terms separately to the two terms in the right hand side, thus obtaining 
sufficient conditions for cancellation of the poles:
\beq\label{d2}
\dot c_i^{\alpha}=\sum_{k=1}^{{\cal N}}M_{ik}c_k^{\alpha}, \qquad
\dot a_i^{\alpha}=-\sum_{k=1}^{{\cal N}}a_k^{\alpha} M_{ki}, \qquad
M_{ik}=2(1-\delta_{ik})\, \frac{{\bf b}_i^T {\bf a}_k}{(x_i-x_k)^2}.
\eeq
Similar calculations with the linear problem for $\Psi^{\dag}$ lead to the same
condition ${\bf b}_i^T {\bf a}_i =1$ and to the equations
\beq\label{d3}
\sum_{k=1}^{{\cal N}}c_k^{*\alpha}(zI-L)_{ki}=a_i^{\alpha}, 
\eeq
\beq\label{d4}
\dot c_i^{*\alpha}=-\sum_{k=1}^{{\cal N}}c_k^{*\alpha}M_{ki}, \qquad
\dot b_i^{\alpha}=\sum_{k=1}^{{\cal N}}M_{ik}b_k^{\alpha}. 
\eeq
Note that the second equations in (\ref{d2}) and (\ref{d4})
give equations of motion for the vectors ${\bf a}_i$ and ${\bf b}_i$.

Therefore, we have the following overdetermined linear problems for the ${\cal N}$-component vectors
${\bf C}^{\alpha}=(c_1^{\alpha}, \ldots , c_{{\cal N}}^{\alpha})^T$ and
${\bf C}^{*\alpha}=(c_1^{*\alpha}, \ldots , c_{{\cal N}}^{*\alpha})^T$:
$$
\left \{
\begin{array}{l}(zI\! -\! L){\bf C}^{\alpha}=-{\bf B}^{\alpha}
\\
\phantom{(zI-L}\dot {\bf C}^{\alpha}=M {\bf C}^{\alpha},
\end{array} \right.
\qquad
\left \{
\begin{array}{l}{\bf C}^{*\alpha T}(zI\! -\! L)={\bf A}^{\alpha}
\\
\phantom{(zI-L}\dot {\bf C}^{*\alpha T}=-{\bf C}^{*\alpha T}M,
\end{array} \right.
$$
where ${\bf A}^{\alpha}=(a_1^{\alpha}, \ldots , a_{{\cal N}}^{\alpha})^T$,
${\bf B}^{\alpha}=(b_1^{\alpha}, \ldots , b_{{\cal N}}^{\alpha})^T$.
The consistency of these linear problems implies (after applying $\p_{t_2}$ to the first one in the first pair):
$$
-\dot L {\bf C}^{\alpha}+(zI-L)\dot {\bf C}^{\alpha}=-\dot {\bf B}^{\alpha}
=-M {\bf B}^{\alpha}=M(zI-L){\bf C}^{\alpha},
$$
i.e., $\Bigl (\dot L-[M,L]\Bigr ){\bf C}^{\alpha}=0$. The second pair of the linear problems
yields, in a similar way, ${\bf C}^{*\alpha T}\Bigl (\dot L-[M,L]\Bigr )=0$. Therefore, the 
consistency condition for the linear problems is
\beq\label{lax}
\dot L=[M,L]
\eeq
which is the Lax equation for our model. 
Using equations of motion for the vectors ${\bf a}_i$ and ${\bf b}_i$, one can check that
non-diagonal parts of the Lax equation are satisfied identically while the diagonal parts
yield equations of motion for the poles $x_i$:
\beq\label{d5}
\ddot x_i=-8\sum_{k\neq i}\frac{({\bf b}_i^T{\bf a}_k)({\bf b}_k^T{\bf a}_i)}{(x_i-x_k)^3}.
\eeq
Together with the equations for the vectors ${\bf a}_i$, ${\bf b}_i$
(see (\ref{d2}), (\ref{d4})),
\beq\label{d6}
\dot {\bf a}_i= -2\sum_{k\neq i}\frac{({\bf b}_k^T {\bf a}_i)\, {\bf a}_k}{(x_i-x_k)^2},
\qquad
\dot {\bf b}_i= 2\sum_{k\neq i}\frac{({\bf b}_i^T {\bf a}_k)\, {\bf b}_k}{(x_i-x_k)^2},
\eeq
they form the closed set of equations of motion for the model. Note that the equations
(\ref{d6}) are compatible with the constraints ${\bf b}_i^T{\bf a}_i=1$. We see that our dynamical
system is the spin generalization of the Calogero system (the Gibbons-Hermsen model).
It is a Hamiltonian system with the Hamiltonian
\beq\label{d7}
H=\sum_{i=1}^{{\cal N}}p_i^2-\sum_{i\neq k}
\frac{({\bf b}_i^T{\bf a}_k)({\bf b}_k^T{\bf a}_i)}{(x_i-x_k)^2}
\eeq
and the non-vanishing Poisson brackets
$
\{x_i, p_k\}=\delta_{ik}, \, \{a_i^{\alpha}, b_k^{\beta}\}=
\delta_{\alpha \beta}\delta_{ik}
$. The Hamiltonian equations of motion
$$
\dot x_i =\frac{\p H}{\p p_i}, \quad \dot p_i =-\frac{\p H}{\p x_i}, \quad
\dot a_i^{\alpha} =\frac{\p H}{\p b_i^{\alpha}}, \quad
\dot b_i^{\alpha} =-\frac{\p H}{\p a_i^{\alpha}}
$$
are equivalent to (\ref{d5}), (\ref{d6}). Taking into account that $\dot x_i =2p_i$, we see that
\beq\label{d8}
H=H_2=\mbox{tr}\, L^2.
\eeq
The spin generalization of the Calogero system is an integrable model. 
The higher Hamiltonians in involution are given by
\beq\label{d9}
H_k=\mbox{tr}\, L^k, \qquad k\geq 1.
\eeq
The Hamiltonian $H_1=-\sum_i p_i$ is the (minus) total momentum.

\subsection{Dynamics in the higher times}

The main tool for investigating the dynamics in higher times is the relation (\ref{m11})
which we write here in the form
\beq\label{d10}
\res_{\infty}\Bigl (z^m \Psi_{\alpha \gamma}\Psi ^{\dag}_{\gamma \beta}\Bigr )
=-\p_{t_m}w_{\alpha \beta}^{(1)}.
\eeq
(We use the notation $\res_{\infty}f(z)=\frac{1}{2\pi i}
\oint_{C_{\infty}}f(z)dz =f_{-1}$, where $f_{-1}$ 
is the coefficient in front of $z^{-1}$ in the Laurent expansion $f(z)=\sum_k f_k z^k$.)
In order to use it, we prepare the following expressions:
$$
\Psi_{\alpha \gamma}\Psi ^{\dag}_{\gamma \beta}=\delta_{\alpha \beta}+
\sum_i \frac{c_i^{*\alpha}b_i^{\beta}}{x-x_i}+\sum_i \frac{a_i^{\alpha}c_i^{\beta}}{x-x_i}
+\sum_{i\neq k}\frac{a_i^{\alpha}c_i^{\gamma}c_k^{*\gamma}b_k^{\beta}+
a_k^{\alpha}c_k^{\gamma}c_i^{*\gamma}b_i^{\beta}}{(x_i-x_k)(x-x_i)}+
\sum_i \frac{a_i^{\alpha}c_i^{\gamma}c_i^{*\gamma}b_i^{\beta}}{(x-x_i)^2},
$$
$$
-\p_{t_m}w_{\alpha \beta}^{(1)}=\sum_i \frac{\p_{t_m}(a_i^{\alpha}b_i^{\beta})}{x-x_i}+
\sum_i \frac{a_i^{\alpha}b_i^{\beta}\, \p_{t_m}x_i}{(x-x_i)^2}.
$$
Comparing the second order poles at $x=x_i$ in (\ref{d10}), we obtain
\beq\label{d11}
\p_{t_m}x_i=\res_{\infty}\Bigl (z^m c_i^{\gamma}c_i^{*\gamma}\Bigr ).
\eeq
Now we solve the linear equations (\ref{d1}), (\ref{d3}) for $c_i^{\gamma}$, $c_i^{*\gamma}$:
\beq\label{d11a}
c_i^{\gamma}=-\sum_k (zI-L)^{-1}_{ik}b_k^{\gamma}, \qquad
c_i^{*\gamma}=\sum_k (zI-L)^{-1}_{ki}a_k^{\gamma}
\eeq
and substitute this into (\ref{d11}). We get:
$$
\p_{t_m}x_i=-\res_{\infty}\Bigl (z^m \sum_{k,l}a_k^{\gamma}(zI-L)^{-1}_{ki}
(zI-L)^{-1}_{il}b_l^{\gamma}\Bigr ).
$$
Recalling that $p_k =\dot x_k/2$ and using the obvious relation 
$\displaystyle{\frac{\p L_{jn}}{\p p_i}=-\delta_{ij}\delta_{in}}$, we can write this as
$$
\p_{t_m}x_i=\res_{\infty}\left (z^m \!\! \sum_{k,l, j, n}a_k^{\gamma}(zI-L)^{-1}_{kj}\,
\frac{\p L_{jn}}{\p p_i}\, (zI-L)^{-1}_{nl}\, b_l^{\gamma}\right )
$$
$$
=\frac{\p}{\p p_i}\, \res_{\infty}\left (z^m  \sum_{k,l}a_k^{\gamma}(zI-L)^{-1}_{kl}\,
b_l^{\gamma}\right )
=\, \frac{\p}{\p p_i}\,\sum_{k,l}(L^m)_{kl}\, {\bf b}_l^T {\bf a}_k
=\frac{\p}{\p p_i}\,\mbox{tr}\Bigl (L^m R\Bigr ),
$$
where $R$ is the ${\cal N}\! \times \! {\cal N}$ matrix with matrix elements $R_{ij}={\bf b}_i^T {\bf a}_j$.
Introduce the  matrix $X=\mbox{diag}\, (x_1, \ldots , x_{{\cal N}})$,
then it is easy to check that
$R=I+[L, X]$. Therefore, $\mbox{tr}\Bigl (L^m R\Bigr ) =\mbox{tr}\Bigl (L^m +L^m(LX-XL)\Bigr )=
\mbox{tr}\, L^m =H_m$ and we obtain one set of the Hamiltonian equations for the higher flow $t_m$:
\beq\label{d12}
\p_{t_m}x_i=\frac{\p H_m}{\p p_i}, \qquad m\geq 2.
\eeq
Note that formally these equations hold also for $m=1$ yielding $\p_{t_1}x_i=-1$ which is true
because of (\ref{m16}). 

In order to obtain the second set of Hamiltonian equations, we apply $\p_{t_2}$ to the both sides
of equation (\ref{d11}):
$$
\p_{t_m}\dot x_i =\res_{\infty}\, \Bigl (z^m 
(\dot c_i^{\gamma}c_i^{*\gamma}+ c_i^{\gamma}\dot c_i^{*\gamma})\Bigr )
=\res_{\infty} \left (z^m \sum_k \Bigl (c_i^{*\gamma}M_{ik}c_k^{\gamma}-c_k^{*\gamma}M_{ki}c_i^{\gamma}\Bigr )
\right )
$$
Next we substitute (\ref{d11a}):
$$
\p_{t_m}\dot x_i =-\res_{\infty}\left (z^m \! \sum_{k,l,n}\Bigl (a_l^{\gamma}
(zI\! -\! L)^{-1}_{li}M_{ik}(zI\! -\! L)^{-1}_{kn}b_n^{\gamma}-
a_l^{\gamma}
(zI\! -\! L)^{-1}_{lk}M_{ki}(zI\! -\! L)^{-1}_{in}b_n^{\gamma}\Bigr ) \right )
$$
$$
\begin{array}{l}
\displaystyle{=-\res_{\infty}\left (z^m \!\! \sum_{k,l,n,j,r}\Bigl (a_l^{\gamma}(zI\! -\! L)^{-1}_{lj}
(E_{ii})_{jr}M_{rk}(zI\! -\! L)^{-1}_{kn}b_n^{\gamma}\right.}
\\ \\
\displaystyle{\left. \phantom{aaaaaaaaaaaaaaaaaa\sum_{k,l,n}} -a_l^{\gamma}
(zI\! -\! L)^{-1}_{lk}M_{kr}(E_{ii})_{rj}(zI\! -\! L)^{-1}_{jn}b_n^{\gamma}\Bigr ) \right )}
\end{array}
$$
$$
=-\res_{\infty}\left (z^m \!\! \sum_{k,l,n,j}a_l^{\gamma}(zI\! -\! L)^{-1}_{lj}\,
[E_{ii}, M]_{jk} (zI\! -\! L)^{-1}_{kn}\, b_n^{\gamma} \right ),
$$
where $E_{ii}$ is the matrix with matrix elements $(E_{ii})_{jr}=\delta_{ij}\delta_{ir}$
($1$ in the place $ii$ and zeros elsewhere). It is easy to check that
$$
[E_{ii}, M]_{jk}=2\frac{\p L_{jk}}{\p x_i}.
$$
Therefore, it holds
$$
\p_{t_m} p_i =-\res_{\infty}\left (z^m \!\! \sum_{k,l,n,j}a_l^{\gamma}(zI\! -\! L)^{-1}_{lj}\,
\frac{\p L_{jk}}{\p x_i}\, (zI\! -\! L)^{-1}_{kn}\, b_n^{\gamma} \right )
$$
$$
=-\frac{\p}{\p x_i} \res_{\infty}\left (z^m \! \sum_{l,n}a_l^{\gamma}(zI\! -\! L)^{-1}_{ln}\,
b_n^{\gamma} \right )
$$
$$
=-\frac{\p}{\p x_i}\sum_{l,n}a_l^{\gamma}(L^m)_{ln}b_n^{\gamma}
=-\frac{\p}{\p x_i}\, \mbox{tr}\, \Bigl (L^m R\Bigr )=-\frac{\p}{\p x_i}\, \mbox{tr}\, \Bigl (L^m \Bigr )
$$
and we obtain the second set of Hamiltonian equations
\beq\label{d13}
\p_{t_m}p_i=-\frac{\p H_m}{\p x_i}, \qquad m\geq 1.
\eeq

Let us turn to the first order poles in (\ref{d10}). Equating the coefficients at the first
order poles in the both sides, we obtain the equation
$$
\p_{t_m}(a_i^{\alpha}b_i^{\beta})=
\res_{\infty}\left (z^m \Bigl (c_i^{*\alpha}b_i^{\beta}+a_i^{\alpha}c_i^{\beta} +\sum_{k\neq i}
\frac{a_i^{\alpha}c_i^{\gamma}c_k^{*\gamma}b_k^{\beta}+
a_k^{\alpha}c_k^{\gamma}c_i^{*\gamma}b_i^{\beta}}{x_i-x_k}
\Bigr )\right )
$$
which can be rewritten as
$$
b_i^{\beta}\left [\p_{t_m}a_i^{\alpha}-\res_{\infty}\left (z^m \Bigl (c_i^{*\alpha}+
\sum_{k\neq i}\frac{a_k^{\alpha}c_k^{\gamma}c_i^{*\gamma}}{x_i-x_k}\Bigr )\right )\right ]
$$
$$
+\,
a_i^{\alpha}\left [\p_{t_m}b_i^{\beta}-\res_{\infty}\left (z^m \Bigl (c_i^{\beta}+
\sum_{k\neq i}\frac{c_i^{\gamma}c_k^{*\gamma}b_k^{\beta}}{x_i-x_k}\Bigr )\right )\right ]=0
$$
Equating to zero expressions in both square brackets separately (which gives sufficient
conditions for cancellation of first order poles), we get a system of evolutionary 
equations for $a_i^{\alpha}$ and $b_i^{\beta}$:
\beq\label{d14}
\p_{t_m}a_i^{\alpha}=\res_{\infty}\left (z^m \Bigl (c_i^{*\alpha}+
\sum_{k\neq i}\frac{a_k^{\alpha}c_k^{\gamma}c_i^{*\gamma}}{x_i-x_k}\Bigr )\right ),
\eeq
\beq\label{d15}
\p_{t_m}b_i^{\beta}=\res_{\infty}\left (z^m \Bigl (c_i^{\beta}+
\sum_{k\neq i}\frac{c_i^{\gamma}c_k^{*\gamma}b_k^{\beta}}{x_i-x_k}\Bigr )\right ).
\eeq
Consider the first equation. Substituting (\ref{d11a}) for $c_i^{\gamma}$, $c_i^{*\gamma}$, we have:
$$
\p_{t_m}a_i^{\alpha}=\res_{\infty}\left [ z^m \left (\sum_k a_k^{\alpha}(zI-L)^{-1}_{ki}-
\sum_{k,l,n}a_l^{\gamma}(zI-L)^{-1}_{li}\, \frac{(1-\delta_{ik})a_k^{\alpha}}{x_i-x_k}\,
(zI-L)^{-1}_{kn}b_n^{\gamma}\right )\right ]
$$
$$
=\res_{\infty}\left [ z^m \left (\sum_k a_k^{\alpha}(zI-L)^{-1}_{ki}+
\sum_{k,l,n,j}a_l^{\gamma}(zI-L)^{-1}_{lj}\, \frac{\p L_{jk}}{\p b_i^{\alpha}}\,
(zI-L)^{-1}_{kn}b_n^{\gamma}\right )\right ]
$$
$$
=\, \frac{\p}{\p b_i^{\alpha}}\res_{\infty}\left (z^m 
\sum_{k,l}a_l^{\gamma}(zI-L)^{-1}_{lk}b_k^{\gamma}\right )
$$
$$
=\, \frac{\p}{\p b_i^{\alpha}}\sum_{l,k}a_l^{\gamma}(L^m)_{lk}b_k^{\gamma}=
\frac{\p}{\p b_i^{\alpha}}\, \mbox{tr}\Bigl (L^m R\Bigr )=\frac{\p}{\p b_i^{\alpha}}\, 
\mbox{tr}\Bigl (L^m \Bigr ),
$$
so we obtain the Hamiltonian equations 
\beq\label{d16}
\p_{t_m}a_i^{\alpha}=\frac{\p H_m}{\p b_i^{\alpha}}, \qquad m\geq 1.
\eeq
In a similar way, from (\ref{d15}) we obtain the Hamiltonian equations
\beq\label{d17}
\p_{t_m}b_i^{\alpha}=-\, \frac{\p H_m}{\p a_i^{\alpha}}, \qquad m\geq 1.
\eeq
Therefore, we have shown that the evolution of $x_i, a_i^{\alpha}, b_i^{\beta}$ 
with respect to the higher times $t_m$ of the matrix KP hierarchy is governed by 
the higher Hamiltonians $H_m$ of the spin Calogero-Moser system. 

\section*{Acknowledgments}

We would like to thank A. Zotov for discussions and comments.
The work of A.Z. was
supported by RSF grant 16-11-10160.

\end{document}